\documentclass[a4paper,11pt]{article}
\usepackage{graphicx,epsfig,picins}
\usepackage{fancyhdr,fancybox}
\usepackage{indentfirst}
\usepackage{verbatim}
\usepackage[sort&compress, numbers]{natbib}
\usepackage{geometry}
\usepackage{extarrows} 
\usepackage[small]{caption2}
\usepackage{extarrows}
\usepackage{chemarrow}
\usepackage{color}
\usepackage{microtype}
\DisableLigatures[f]{encoding = *, family = *}

\usepackage{times}     

\usepackage{amssymb}                 
\usepackage{amsthm}
\usepackage{bbm}
\usepackage{hyperref}  

\newcommand{\paperfont}{\fontsize{12pt}{1.3\baselineskip}\selectfont}
\begin{document}

\theoremstyle{definition}
\makeatletter
\thm@headfont{\bf}
\makeatother
\newtheorem{definition}{Definition}
\newtheorem{example}{Example}
\newtheorem{theorem}{Theorem}
\newtheorem{lemma}{Lemma}
\newtheorem{corollary}{Corollary}
\newtheorem{remark}{Remark}
\newtheorem{proposition}{Proposition}

\lhead{}
\rhead{}
\lfoot{}
\rfoot{}

\renewcommand{\refname}{References}
\renewcommand{\figurename}{Figure}
\renewcommand{\tablename}{Table}
\renewcommand{\proofname}{Proof}

\newcommand{\diag}{\mathrm{diag}}
\newcommand{\one}{\mathbbm{1}}


\title{\textbf{Stochastic fluctuations can reveal the feedback signs of gene regulatory networks at the single-molecule level}}
\author{Chen Jia$^1$,\;\;Peng Xie$^{2}$,\;\;Min Chen$^{1,*}$,\;\;Michael Q. Zhang$^{2,3,*}$ \\
\footnotesize $^1$Department of Mathematical Sciences, University of Texas at Dallas, Richardson, TX 75080, U.S.A. \\
\footnotesize $^2$Department of Biological Sciences, Center for Systems Biology, University of Texas at Dallas, Richardson, TX 75080, U.S.A. \\
\footnotesize $^3$MOE Key Lab and Division of Bioinformatics, CSSB, TNLIST, Tsinghua University, Beijing 100084, China\\
\footnotesize $^*$Correspondence: mchen@utdallas.edu (M.C.), michael.zhang@utdallas.edu (M.Q.Z).}
\date{}                              
\maketitle                           
\thispagestyle{empty}                

\paperfont

\begin{abstract}
Understanding the relationship between spontaneous stochastic fluctuations and the topology of the underlying gene regulatory network is of fundamental importance for the study of single-cell stochastic gene expression. Here by solving the analytical steady-state distribution of the protein copy number in a general kinetic model of stochastic gene expression with nonlinear feedback regulation, we reveal the relationship between stochastic fluctuations and feedback topology at the single-molecule level, which provides novel insights into how and to what extent a feedback loop can enhance or suppress molecular fluctuations. Based on such relationship, we also develop an effective method to extract the topological information of a gene regulatory network from single-cell gene expression data. The theory is demonstrated by numerical simulations and, more importantly, validated quantitatively by single-cell data analysis of a synthetic gene circuit integrated in human kidney cells. \\

\noindent 
\end{abstract}

\section*{Introduction}
\begin{figure}[!htb]
\centering{}\centerline{\includegraphics[width=0.6\textwidth]{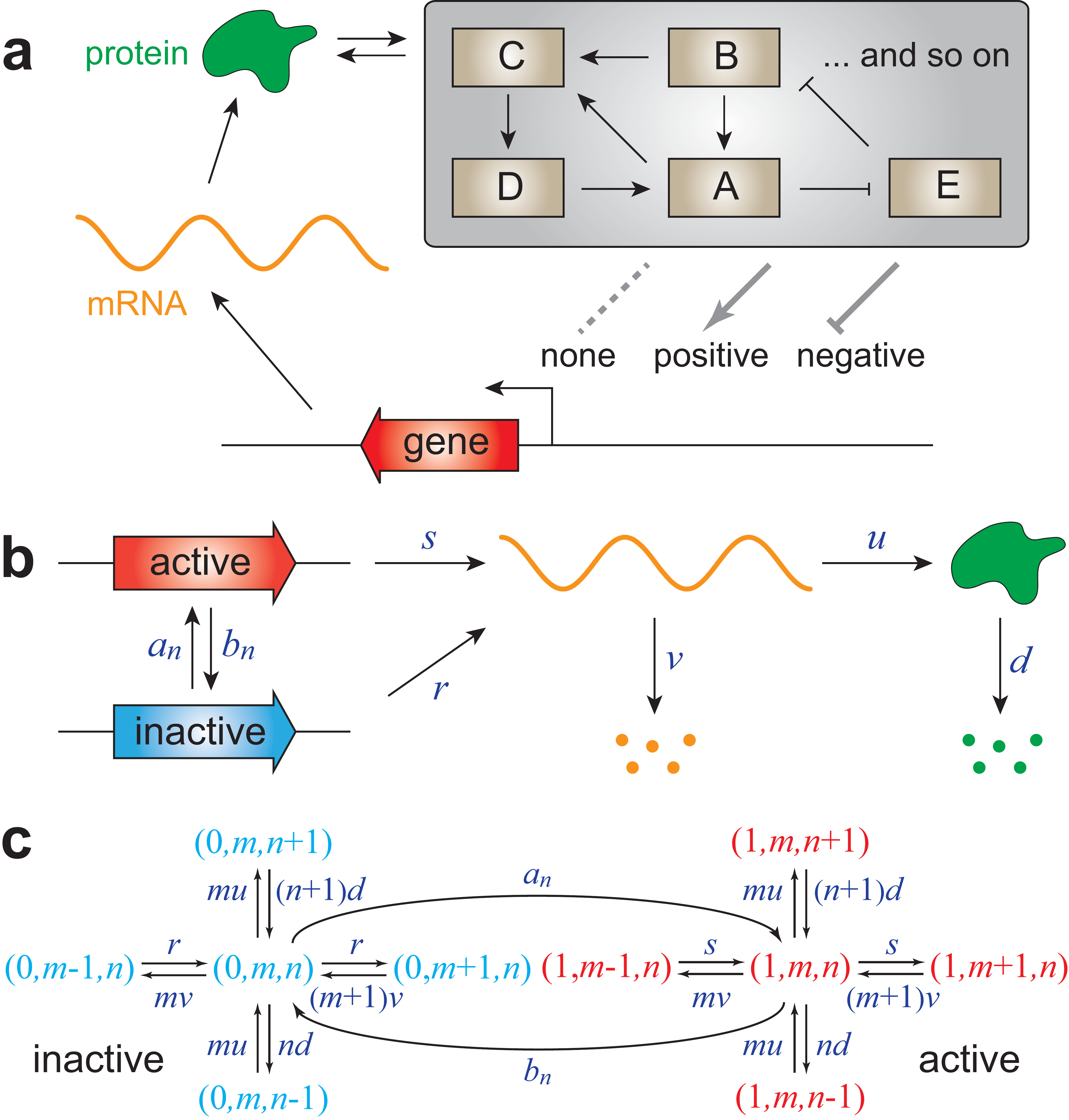}}
\caption{\textbf{Schematic diagrams of stochastic gene expression in living
cells.}
\textbf{(a)} Three types of fundamental autoregulatory topological structures. Gene regulatory networks in a living cell can be overwhelmingly complex, involving numerous feedback loops and signaling steps. However, if one focuses on a particular gene of interest (red), then there are only three types of fundamental regulatory relations: no feedback (none), positive feedback, and negative feedback. The dotted line denotes that there is no link between adjacent nodes.
\textbf{(b)} Three-stage model of stochastic gene expression. The promoter of interest can transition between an active and an inactive epigenetic forms. Since the network has feedback regulation, the switching rates of the promoter depend on the protein copy number.
\textbf{(c)} Markov dynamics associated with the three-stage model. The biochemical state of the gene can be represented by three variables: the promoter activity $i$, the mRNA copy number $m$, and the protein copy number $n$.}
\label{network}
\end{figure}

Gene expression in living cells is a complex stochastic process characterized by various probabilistic chemical reactions, giving rise to spontaneous fluctuations in the abundances of proteins and mRNAs \cite{raj2008nature, xie2008single, eldar2010functional, sanchez2013regulation}. Recent advances in experiment techniques, such as flow cytometry, fluorescence microscopy, and scRNA-Seq, have resulted in the generation of large amounts of single-cell gene expression data. This raises a great challenge of whether and how one can infer the topological structure of a gene regulatory network by using such massive but often noisy data. Considering the complexity of gene regulatory networks, this may seem to be a daunting task. However, the situation becomes much simpler if we focus on a particular gene of interest and the feedback loop regulating it \cite{lestas2010fundamental}. In general, there are only three types of gross topological structures: no feedback, positive feedback, and negative feedback (see Fig. \ref{network}a) and different types of networks can give rise to \emph{similarly shaped}, usually unimodal, steady-state distributions of gene expression. Therefore, it is highly nontrivial to ask whether the information of feedback topology can be extracted from single-cell measurements of this gene.

\section*{Results}

\subsection*{Model and steady-state protein distribution}
Recently, significant progress has been made in the field of single-cell stochastic gene expression \cite{peccoud1995markovian, paulsson2000random, kepler2001stochasticity, hornos2005self, friedman2006linking, raj2006stochastic, shahrezaei2008analytical, assaf2011determining, grima2012steady, kumar2014exact, jia2015modeling, ge2015stochastic, pajaro2015shaping, lin2016gene, liu2016decomposition, jia2017emergent}. Based on the central dogma of molecular biology, the kinetics of stochastic gene expression in a single cell can be described by a model with three stages consisting of transcription, translation, and switching of the promoter between an active and an inactive epigenetic forms (see Fig. \ref{network}b). This model is similar to the three-stage model introduced in \cite{shahrezaei2008analytical} but with a critical addition of nonlinear feedback regulation. The biochemical state of the gene of interest can be described by three variables: the activity $i$ of its promoter with $i=1$ and $i=0$ corresponding to the active and inactive forms, respectively, the copy number $m$ of the mRNA transcript, and the copy number $n$ of the protein product. The evolution of the three-stage model can be mathematically described by the Markov dynamics illustrated in Fig. \ref{network}c. Here $s$ and $r$ are the transcription rates when the promoter is active and inactive, respectively (the basal transcription rate $r$ is usually not zero), $u$ is the translation rate, and $v$ and $d$ are the degradation rates of the mRNA and protein, respectively. Since the network has feedback regulation, the protein copy number $n$ will directly or indirectly affect the switching rates $a_n$ and $b_n$ of the promoter between the active and inactive forms. Since many genes have complex epigenetic controls including dissociation of repressors, association of activators, or chromatin remodeling, we do not impose any restrictions on the specific functional forms of $a_n$ and $b_n$. In \cite{kumar2014exact}, the authors considered the case of linear feedback regulation with $a_n = a+un$ and $b_n = b$, where $a$ is the spontaneous contribution and $un$ is the feedback contribution with $u$ measuring the feedback strength. However, recent single-cell experiments on transcription of mammalian cells \cite{bintu2016dynamics} suggest that $a_n$ and $b_n$ are often saturated when $n\gg 1$ and thus are highly nonlinear. In the present work, we consider a more general case by allowing arbitrary nonlinearity.

In most applications, the switching rates of the promoter are fast \cite{friedman2006linking, ge2015stochastic} and the effective transcription rate of the gene is given by $c_n=(a_ns+b_nr)/(a_n+b_n)$. It is critical to note that the information of network topology is implicitly characterized by  $c_n$. If the network has a positive-feedback (negative-feedback) loop, then $c_n$ is an increasing (decreasing) function of $n$. If the network has no feedback, $c_n$ is independent of $n$. Let $p_n$ denote the steady-state probability of having $n$ protein molecules. Experimentally, the lifetime of the mRNA is usually much shorter compared to that of its protein counterpart \cite{shahrezaei2008analytical}. Once an mRNA is synthesized, it can either produce a protein with probability $p = u/(u+v)$ or degrade with probability $q = v/(u+v)$. Let $\lambda = v/d$ denote the ratio of the protein and mRNA lifetimes. When $\lambda\gg1$, the original Markov model can be simplified to a reduced model with geometrically distributed translation bursts \cite{jia2017simplification} and the steady-state distribution of the protein copy number can be calculated analytically (Supplementary Information):
\begin{equation}\label{distribution}
p_n = A\frac{p^n}{n!}
\frac{c_{0}}{d}\left(\frac{c_{1}}{d}+1\right)\cdots\left(\frac{c_{n-1}}{d}+n-1\right),
\end{equation}
where $A$ is a normalization constant. If the network has no feedback, then $c_n = c$ is a constant and the above distribution reduces to the well-known negative-binomial distribution
\begin{equation*}
p_n = \frac{p^n}{n!}\frac{\Gamma(c/d+n)}{\Gamma(c/d)}q^{c/d},
\end{equation*}
where $\Gamma(x)$ is the gamma function. This is consistent with the results obtained in \cite{paulsson2000random, shahrezaei2008analytical}.

In fact, the parameter $q$ has important statistical implications. Since $c_n\leq s$, it follows from Eq. \eqref{distribution} that $p_{n+1}/p_n = p(n+c_n/d)/(n+1)\approx p$ when $n\gg1$. This further suggests that $p_{n+k}\approx p^{k}p_n = e^{k\log(1-q)}p_n\approx e^{-qk}p_n$ when $q\ll p$. This shows that the steady-state probability $p_n$ decays exponentially with respect to the protein copy number $n$ when $n\gg1$ with $q$ being the exponentially decaying rate of the steady-state protein distribution\footnote{Here the decaying rate is a mathematical concept, rather than a physical or chemical concept such as a chemical reaction rate.}. Here $q\ll p$ is justified because $p/q = u/v$ is the average number of proteins synthesized per mRNA lifetime, which is relatively large in living cells and typically on the order of 100 for an \emph{E. coli} gene \cite{paulsson2005models}. To identify $q$ as an experimentally accessible quantity is of basic importance, as will be shown later.

\subsection*{Decomposition of the protein fluctuations}
Experimentally, spontaneous stochastic fluctuations, often referred to as noise, in the protein abundance are usually measured by the squared relative standard deviation $\eta = \sigma^2/\langle n\rangle^2$, where $\langle n\rangle$ is the mean and $\sigma^2$ is the variance \cite{paulsson2004summing}. With the analytical steady-state protein distribution, it can be shown that the noise $\eta$ can be decomposed into three different terms or two different terms as (Supplementary Information)
\begin{eqnarray}
\eta & = & \frac{1}{\langle n\rangle}+\frac{d}{v\langle m\rangle}+\eta_f\begin{cases}
\eta_f=0\;\;\;\text{no feedback}\\
\eta_f>0\;\;\;\text{positive feedback}\\
\eta_f<0\;\;\;\text{negative feedback}
\end{cases}\label{noisenumber}\\
 & = & \frac{1}{q\langle n\rangle}+\eta_f,\label{noise}
\end{eqnarray}
where $1/\langle n\rangle$ is the Poisson noise from individual births and deaths of the protein, $d/v\langle m\rangle$ is the noise due to fluctuations in the mRNA abundance, and $\eta_f = \textrm{Cov}(n,c_n)/\langle n\rangle\langle c_n\rangle$ is the relative covariance between $n$ and $c_n$, which characterizes the strength of feedback regulation. We stress here that when the promoter switching rates are fast, the above decomposition formula and the expression of $\eta_f$ hold exactly without any approximation, even when the nonlinearity of feedback regulation is very high.

If the network has no feedback, then $c_n$ is a constant and $\eta_f = 0$. It is well know that the covariance between a random variable and an increasing (decreasing) function of this random variable must be positive (negative). Therefore, if the network has a positive-feedback loop, then $c_n$ is an increasing function of $n$ and $\eta_f > 0$. Conversely, if the network has a negative-feedback loop, then $c_n$ is a decreasing function of $n$ and $\eta_f < 0$. Therefore, the sign of $\eta_f$ is completely determined by the network topology and we shall name $\eta_f$ as \emph{feedback coefficient}. The above analysis clearly explains previous experimental observations that positive feedback generally amplifies noise \cite{becskei2001positive} and negative feedback generally reduces noise \cite{becskei2000engineering}.

In the previous literature, there are confusing or even contradictory statements about the feedback-noise relationship. Some studies claimed that positive feedback reduces noise \cite{stekel2008strong}, while negative feedback amplifies noise \cite{hornung2008noise}. The reason for these seemingly contradictory results has been analyzed in \cite{kumar2014exact, liu2016decomposition} and here we shall use our noise decomposition formula to provide an clearer explanation. For a positive-feedback (negative-feedback) network, $\eta$ is the total noise and $\pm\eta_f$ is the noise amplified (reduced). Therefore, $\eta-\eta_f = 1/q\langle n\rangle$ can be thought of as the \emph{feedback-free noise}. In general, if all the other rate constants remain unchanged, then positive (negative) feedback will lead to an increase (decrease) in the protein mean $\langle n\rangle$ \cite{friedman2006linking} and thus lead to a decrease (increase) in the feedback-free noise $1/q\langle n\rangle$. This decrease (increase) in the feedback-free noise may counteract the positive (negative) contribution of the feedback coefficient $\eta_f$ and give rise to an anomalous decrease (increase) in the total noise $\eta$. This explains why some experiments have observed anomalous noise suppression (amplification) in networks with positive (negative) feedback.

However, from the physical perspective, the feedback-free noise and feedback coefficient have completely different origins: the former characterizes fluctuations from individual births and deaths of the protein and mRNA, while the latter reflects the contribution of feedback regulation. Therefore, it seems logically insufficient to study the effect of feedback regulation on the feedback-free noise by fixing the underlying biochemical rate constants. In fact, what positive (negative) feedback actually amplifies (reduces) is the very part of fluctuations that cannot be explained by the feedback-free noise.

\subsection*{Bounds for the protein noise}
Negative feedback proves to be most interesting because it is responsible for the stability of a cell \cite{becskei2000engineering}. Since negative feedback reduces noise, it is natural to ask to what extent the noise is inevitable and whether the feedback coefficient $\eta_f$ could be strong enough such that the noise $\eta$ is approaching zero \cite{lestas2010fundamental, hilfinger2016constraints}. In fact, for the three-stage model, the upper and lower bounds of the noise $\eta$ are given by
\begin{equation}\label{lowerbound}
\frac{1}{q\langle n\rangle}\frac{1}{1+\alpha p/dq} \leq \eta < \frac{1}{q\langle n\rangle},
\end{equation}
where $\alpha=\sup\{|c'(x)|:x>0\}$ is the steepness of the regulatory function $c(x)$ obtained from $c_n$ by replacing $n$ with a positive real number $x$ and the term $\alpha p/dq$ is of the order of one for a wide range of biologically relevant parameters (Supplementary Information). These bounds provide the limits on the ability for a negative-feedback loop to suppress protein fluctuations. We stress here that this lower bound is new and is different from the one derived in \cite{hilfinger2016constraints}. Our lower bound performs better in the regime of strong noise suppression (Supplementary Information). In the literature, the effective transcription rate $c(x)$ is often chosen as the generalized Hill function $c(x) = (as+x^{h}r)/(a+x^{h})$ with $h\geq1$ being the Hill coefficient \cite{friedman2006linking, lestas2010fundamental}, in which case the steepness
\begin{equation*}
\alpha = \frac{(h-1)^{1-1/h}(h+1)^{1+1/h}}{4h}\times\frac{(s-r)}{a^{1/h}}.
\end{equation*}

For a negative-feedback network, $\eta-\eta_f$ is the feedback-free noise, $-\eta_f$ is the noise reduced, and $\eta$ is the total noise. Then the efficiency of the negative-feedback network, as a noise filter, can be defined as $\gamma = -\eta_f/(\eta-\eta_f)$. The lower bound in Eq. \eqref{lowerbound} reveals a general biophysical principle: The efficiency of a negative-feedback network must satisfy $0 < \gamma \leq 1/(1+dq/\alpha p)$. This fact is similar to Carnot's theorem in classical thermodynamics, which claims that the theoretical maximum efficiency of any heat engine must be smaller than 1.

\begin{figure}[!htb]
\centering{}\centerline{\includegraphics[width=0.8\textwidth]{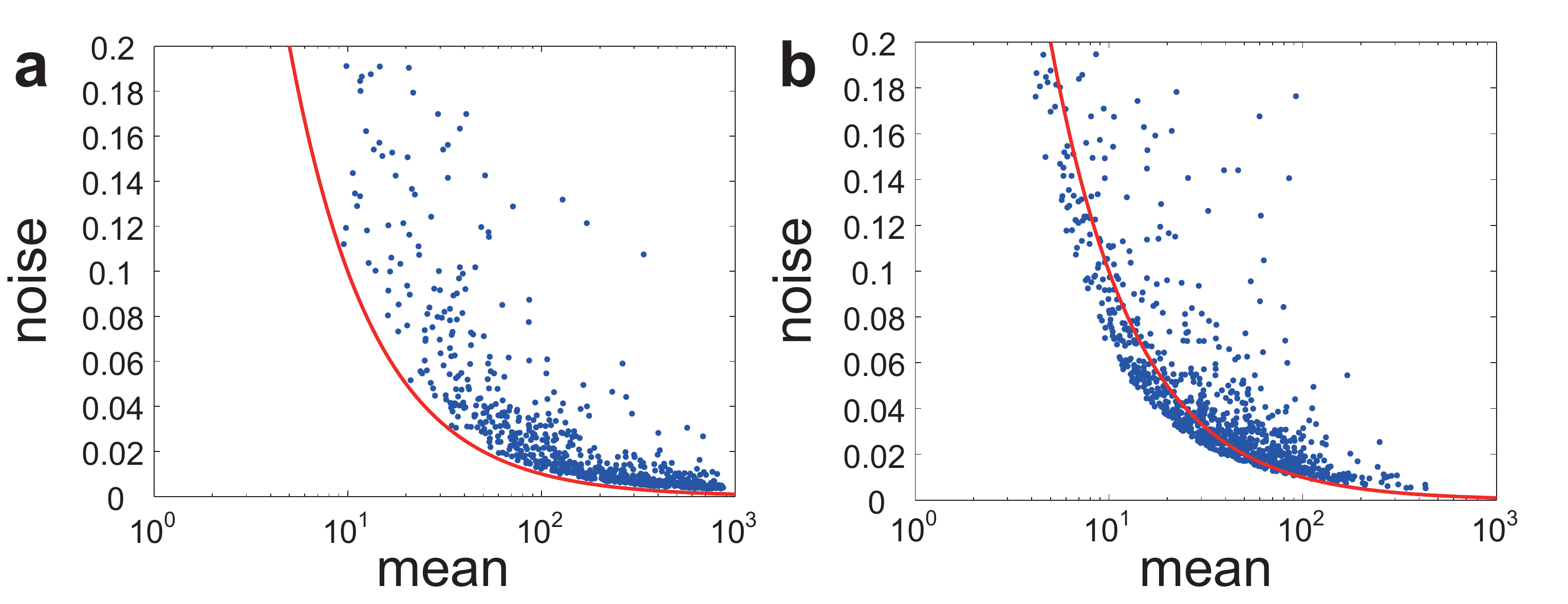}}
\caption{\textbf{Effect of feedback regulation on the protein noise by numerical simulations.}
\textbf{(a)} The protein noise $\eta$ versus the protein mean $\langle n\rangle$ in positive-feedback networks under different choices of model parameters. The functional forms of $a_n$ and $b_n$ are chosen as $a_n = an$ and $b_n = b$.
\textbf{(b)} The protein noise $\eta$ versus the protein mean $\langle n\rangle$ in negative-feedback networks under different choices of model parameters. The functional forms of $a_n$ and $b_n$ are chosen as $a_n = a$ and $b_n = bn$. In both (a) and (b), the red curve represents the Poisson noise and the model parameters are randomly chosen as $s\sim U[10,500],\;r\sim U[0,10],\;d = 1,\;p\sim U[0,1],\;a\sim U[0,1000],\;b\sim U[0,1000]$, where $U[x,y]$ denotes the uniform distribution on the interval $[a,b]$.}\label{Poisson}
\end{figure}

If all other cellular factors are constant, the protein will display a small-number Poisson noise \cite{paulsson2005models}. When $\alpha > d$, the lower bound in Eq. \eqref{lowerbound} is smaller than $1/\langle n\rangle$, which shows that $\eta$ may be even smaller than the Poisson noise in the negative-feedback case (see Fig. \ref{Poisson}b). Recent experiments have shown that although the variance of expression levels is larger than the mean for most genes, there are still some genes whose variance is less than the mean \cite{anders2010differential}. This fact is well explained by our theory. From Eq. \eqref{noise}, if the network has no feedback or a positive-feedback loop, $\eta$ is always larger than the Poisson noise (see Fig. \ref{Poisson}a). In the positive-feedback case, similar upper and lower bounds for the noise $\eta$ can also be obtained (Supplementary Information), which provide the limits on the ability for a positive-feedback loop to enhance protein fluctuations.

\subsection*{Inference of feedback topology using single-cell data}
When a network has nonlinear feedback regulation, the mean and variance are not enough to determine the steady-state protein distribution and the information of higher-order moments will play a crucial role. In fact, Eq. \eqref{noise} can be rewritten in a more illuminating form as
\begin{equation}\label{relation}
\eta_f = \frac{\sigma^2}{\langle n\rangle^2}-\frac{1}{q\langle n\rangle}.
\end{equation}
This equation is of crucial importance because it bridges the feedback topology of a gene circuit and experimentally accessible measurements. In particular, it reveals a quantitative relationship between the feedback coefficient $\eta_f$, whose sign is fully determined by the network topology, and the digital features of the steady-state protein distribution, characterized by the mean $\langle n\rangle$, variance $\sigma^2$, and decaying rate $q$, which reflects the overall effect of higher-order moments.  This provides an effective method to extract the topological information of a gene regulatory network from single-cell gene expression data. From single-cell data, the three digital features, and thus the feedback coefficient $\eta_f$, can be estimated robustly (Supplementary Information). If $\eta_f$ is significantly larger (smaller) than zero, one has good reasons to believe that there is a positive-feedback (negative-feedback) loop regulating this gene.

In single-cell experiments such as flow cytometry and fluorescence microscopy, one usually obtains data of protein concentrations, instead of protein copy numbers. Let $x = n/V$ be a continuous variable representing the protein concentration, where $V$ is a constant compatible with the macroscopic scale. It is easy to see that the noise $\eta = \sigma^2/\langle n\rangle^2$ will not be affected by the scaling constant $V$ and thus is dimensionless. In terms of the protein concentration, the mean will become $\langle n\rangle/V$ and the decaying rate will become $qV$ (Supplementary Information). Therefore, the product of these two terms is also dimensionless. This indicates that the above method not only applies to single-molecule data of protein copy numbers, but also applies to single-cell data of protein concentrations. The above analysis also suggests a crucial difference between the two decomposition formulas \eqref{noisenumber} and \eqref{noise}: The former only applies to data of protein copy numbers, while the latter also applies to data of protein concentrations.

\subsection*{Experimental validation}
\begin{figure}[!htb]
\centering{}\centerline{\includegraphics[width=1\textwidth]{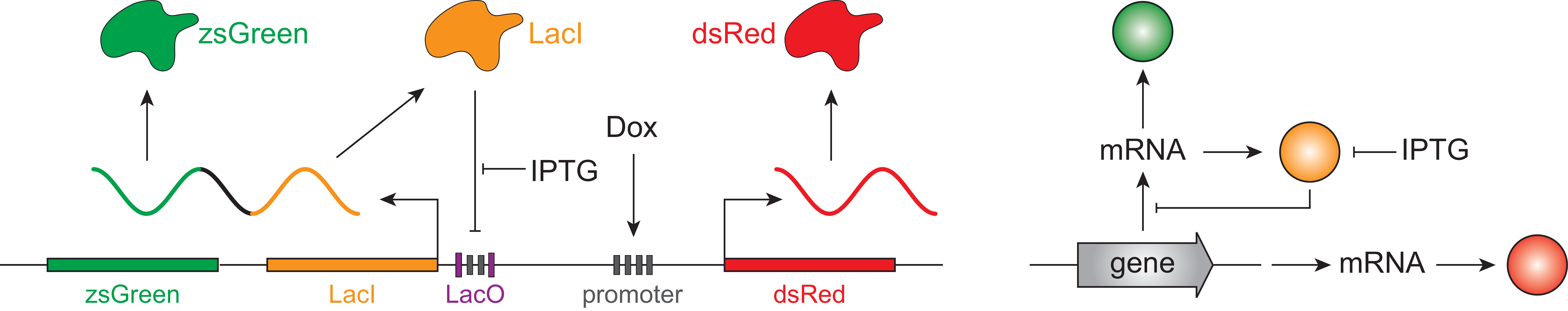}}
\caption{\textbf{A synthetic gene network integrated in human kidney cells.}
The bidirectional promoter transcribes the zsGreen-LacI and dsRed transcripts. The gene network includes two architectures: a negative-feedback network and a network with no feedback. The zsGreen-LacI transcripts
are inhibited by LacI, forming a network with negative autoregulation. The dsRed transcripts are not regulated, forming a network with no feedback.}
\label{circuit}
\end{figure}

To validate our theory, we apply it to a synthetic gene circuit (orthogonal property of a synthetic network can minimize ``extrinsic" noise) stably integrated in human kidney cells, as illustrated in Fig. \ref{circuit}) \cite{shimoga2013synthetic}. In this circuit, a bidirectional promoter is designed to control the expression of two fluorescent proteins: zsGreen and dsRed. The activity of the promoter can be activated in the presence of Doxycycline (Dox). The green fluorescent protein, zsGreen, is fused upstream from the transcriptional repressor LacI. The LacI protein binds to its own gene and inhibits the transcription of its own mRNA, forming a negative-feedback loop. The negative-feedback strength can be tuned by induction of Isopropyl $\beta$-D-1-thiogalactopyranoside (IPTG). As the control architecture, the red fluorescent protein, dsRed, is not regulated by IPTG induction, forming a network with no feedback. The steady-state levels of the zsGreen and dsRed fluorescence are measured under a wide range of IPTG concentrations and two Dox concentrations (low and high) by using flow cytometry.

\begin{figure*}[!htb]
\centerline{\includegraphics[width=0.8\textwidth]{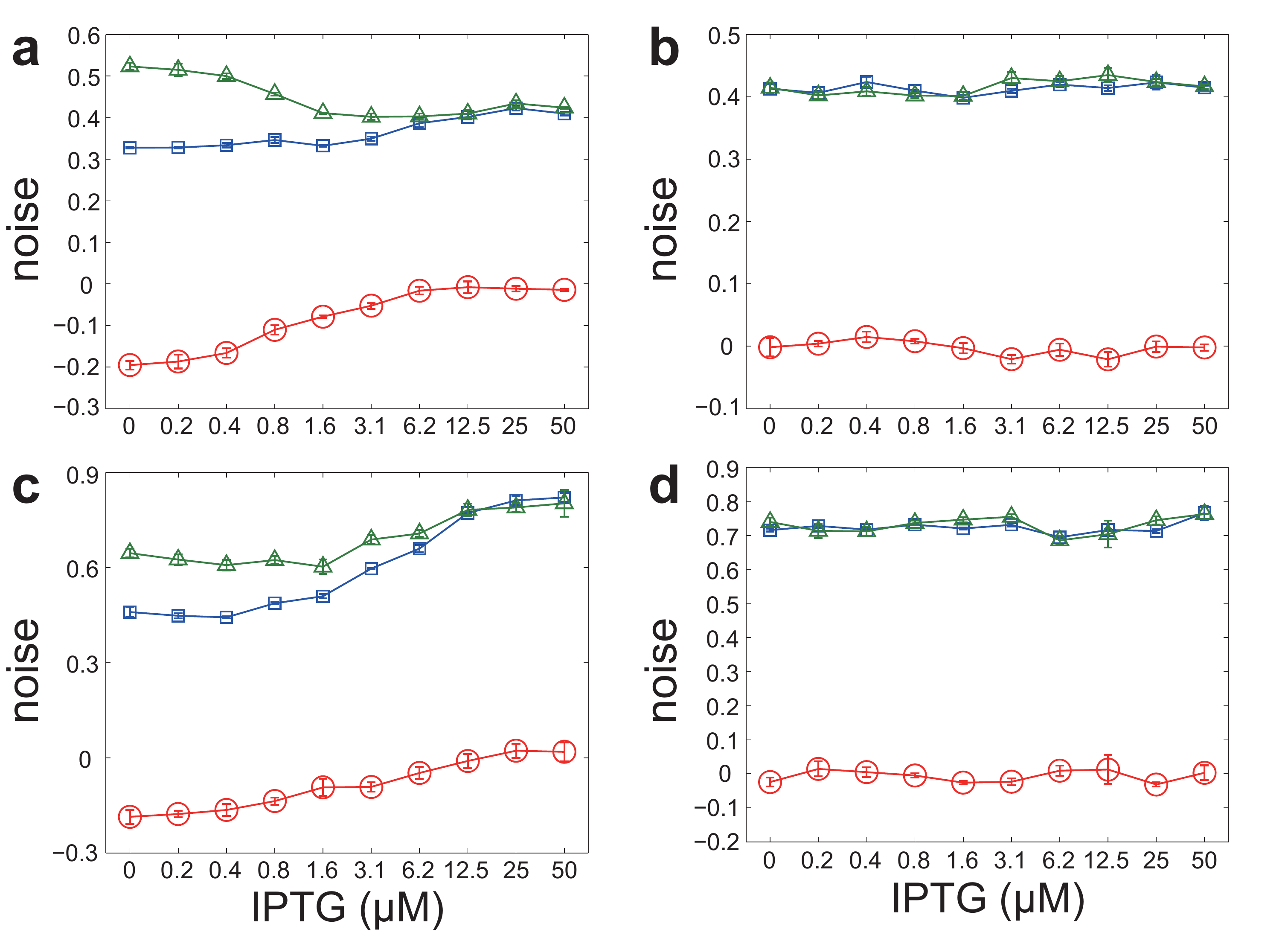}}
\caption{\textbf{Inference of the network topology by using single-cell data.}
\textbf{(a)-(d)} The noise $\eta$ (blue), feedback-free noise $\eta-\eta_f$ (green), and feedback coefficient $\eta_f$ (red) for the zsGreen and dsRed proteins under different IPTG concentrations.
The error bars are standard deviations given by bootstrap.
\textbf{(a)} zsGreen in the high Dox case.
\textbf{(b)} reRed in the high Dox case.
\textbf{(c)} zsGreen in the low Dox case.
\textbf{(d)} reRed in the low Dox case.}\label{inference}
\end{figure*}

For each fixed IPTG and Dox concentrations, we can estimate the mean $\langle n\rangle$, variance $\sigma^2$, and decaying rate $q$ for the steady-state distribution of the zsGreen or dsRed fluorescence. Then the feedback coefficient $\eta_f$ can be estimated from Eq. \eqref{relation}. In the high Dox case, Figs. \ref{inference}a and \ref{inference}b illustrate the noise $\eta$, feedback-free noise $\eta-\eta_f$, and feedback coefficient $\eta_f$ of the zsGreen and dsRed proteins under different IPTG concentrations, respectively. For the zsGreen protein, the feedback coefficient $\eta_f$ is negative under all IPTG concentrations. With the increase of the IPTG concentration, the negative-feedback strength becomes increasingly weaker and the feedback coefficient $\eta_f$ tends to zero. In contrast, for the dsRed protein, the feedback coefficient $\eta_f$ fluctuates around zero in a narrow range under different IPTG concentrations. These results are in full agreement with our theory with high accuracy. As a result, our method correctly extracts the topological information of the synthetic gene circuit in both qualitative and quantitative ways. In the low Dox case, the noise $\eta$, feedback-free noise $\eta-\eta_f$, and feedback coefficient $\eta_f$ of the zsGreen and dsRed proteins are illustrated in Figs. \ref{inference}c and \ref{inference}d, respectively, and similar conclusions can be drawn.

Although it has been observed that negative feedback suppresses molecular fluctuations \cite{shimoga2013synthetic}, it remains difficult to quantify the corresponding effect \cite{lestas2010fundamental}. Our theory provides a quantitative characterization of such effect. In the high Dox case, the negative-feedback effect is the strongest when the IPTG concentration is zero. In this situation, the feedback-free noise is $\eta-\eta_f=$ 0.49 and the feedback coefficient is $\eta_f = -$0.18, which indicates that negative feedback reduces noise by 36.7\%. The efficiency $\gamma$ of the negative-feedback network drops significantly with the increase of the IPTG concentration and is close to zero when the concentration reaches 6.2 $\mu$M.

One of the potential applications of our theory is to provide a mechanism-driven method to identify the differentially expressed genes (DEGs) of two different cell populations such as tumor and non-tumor tissues. Most of the existing methods searched the DEGs by identifying the difference in the mean levels of the two cell populations under some \emph{a priori} assumptions on the protein or mRNA distribution such as the negative binomial distribution \cite{anders2010differential}. However, the effect of noise amplification or suppression caused by feedback loops is not addressed by these methods, which may result in incorrect predictions (Supplementary Information). Our theory indicates that even if the means and variances of the two cell populations are both very close, one is still able to find the DEGs by detecting the difference in feedback topology. If the signs of the estimated feedback coefficients $\eta_f$ of the two cell populations are different, one has good reasons to believe that there is a change in the topological structure of the underlying gene regulatory network when a non-tumor tissue becomes a tumor one.

\section*{Discussion and conclusions}
Here we present a comprehensive analysis of the three-stage model of stochastic gene expression with nonlinear feedback regulation. By taking the limit of a large ratio of protein to mRNA lifetimes, we derive the analytical steady-state distribution of the protein copy number. Furthermore, we decompose the protein noise according to different biophysical origins. The resulting decomposition formula reveals a quantitative relation between stochastic fluctuations and feedback topology at the single-molecule level. In particular, we show that the protein noise $\eta$ can be decomposed into the sum of two parts: the feedback-free noise $1/q\langle n\rangle$ and feedback coefficient $\eta_f$, whose sign is totally determined by the network topology. Both the two parts can be estimated robustly from single-cell gene expression data via three experimentally accessible quantities: the mean $\langle n\rangle$, variance $\sigma^2$, and decaying rate $q$. Such relation not only enables us to quantify the effects of noise amplification or suppression caused by feedback loops, but also allows us to extract the topological information of the underlying gene regulatory network from single-cell gene expression data. The feasibility of this approach is validated quantitatively by single-cell data analysis of a synthetic gene circuit integrated in human kidney cells.

\begin{figure*}[!htb]
\centerline{\includegraphics[width=0.8\textwidth]{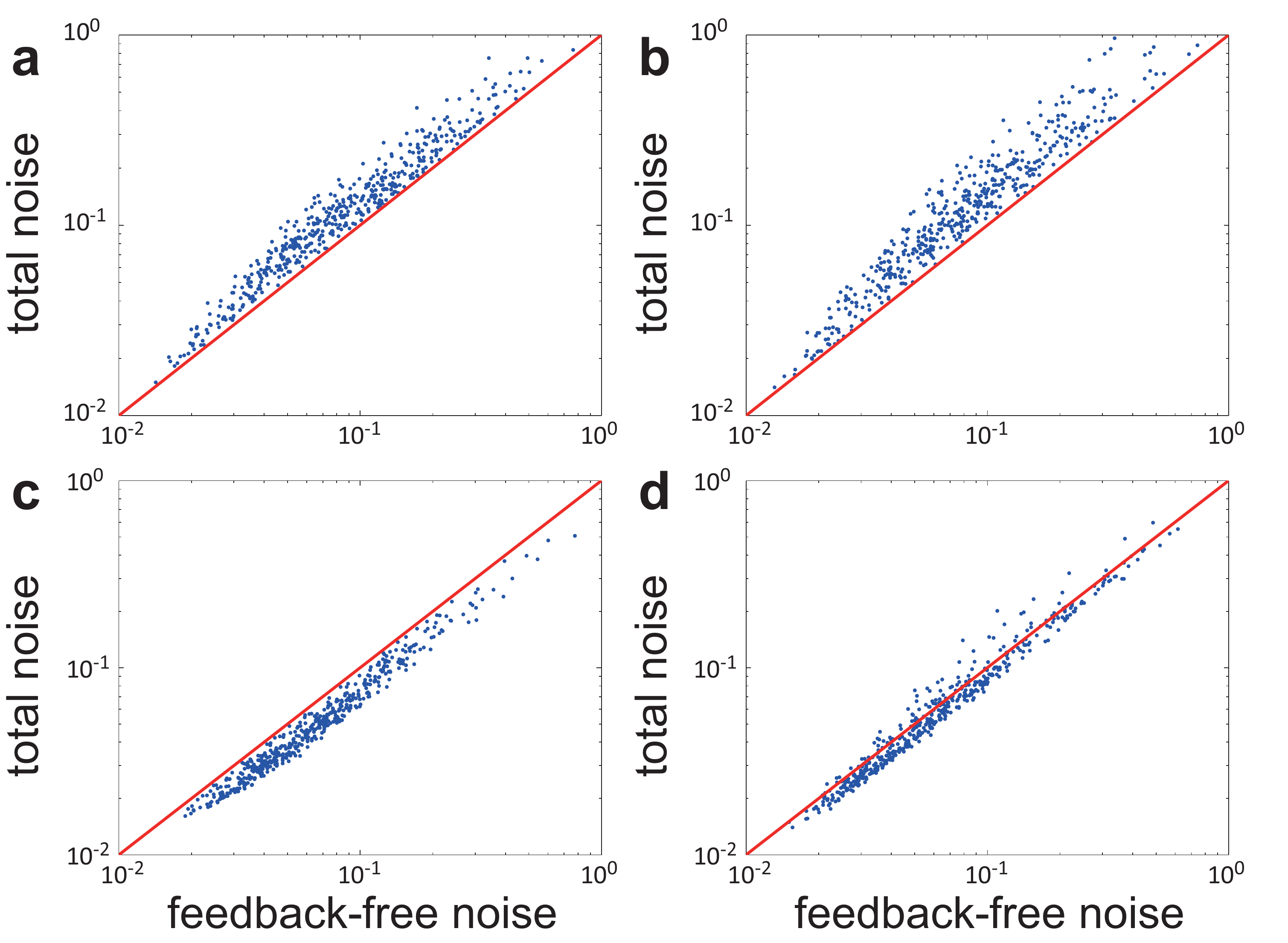}}
\caption{\textbf{Numerical simulations of the total noise versus the feedback-free noise in networks with positive or negative feedback under different choices of model parameters.}
\textbf{(a)} Positive feedback with fast promoter switching.
\textbf{(b)} Positive feedback with slow promoter switching.
\textbf{(c)} Negative feedback with fast promoter switching.
\textbf{(d)} Negative feedback with slow promoter switching.
In (a)-(d), the model parameters are chosen as $r\sim U[0,10]$, $s\sim U[10,100]$, $d = 1$, $v = 30d$, $p\sim U[0.1,0,9]$, $q = 1-p$, where $U[x,y]$ denotes the uniform distribution on the interval $[x,y]$. Since $s$ is the mRNA synthesis rate when the promoter is active and $p/q$ is the average number of proteins synthesized per mRNA lifetime, the maximal protein synthesis rate is given by $s_{\max} = sp/q$. The functional forms of the promoter switching rates are chosen as $a_n = a+un, b_n = b$ in (a),(b) and $a_n = a, b_n = b+un$ in (c),(d). Here the parameters $a,b,u$ are chosen as $a,b\sim U[s_{\max},50s_{\max}], u\sim U[1,50]$ in (a),(c) and $a,b\sim U[0,s_{\max}], u\sim U[0,1]$ in (b),(d).}\label{simulations}
\end{figure*}

We stress that our results depend nothing on the specific functional forms of the effective transcription rate $c_n$ except for its monotonicity, which makes our theory highly general. One of the most powerful parts of our theory is that it can be applied to gene regulatory networks with highly nonlinear feedback. In the present paper, all the derivations are based on the assumption of rapid promoter switching, under which the fluctuations due to promoter switching are averaged out. Intuitively, in the regime of slow promoter switching, our noise decomposition formula \eqref{noise} should be amended as
\begin{equation}\label{full}
\eta = \frac{1}{q\langle n\rangle}+\eta_f+\eta_s,
\end{equation}
where $\eta_s>0$ is the noise due to promoter switching. Because of the contribution of $\eta_s$, the difference between the total noise $\eta$ and feedback-free noise $1/q\langle n\rangle$ must be positive in positive-feedback networks and may be either positive or negative in negative-feedback networks due to the competition of $\eta_f<0$ and $\eta_s>0$. The above analysis is in full agreement with our numerical simulations in Fig. \ref{simulations}. The ignorance of $\eta_s$ in the present paper is the cost for deriving an analytical protein distribution in networks with nonlinear feedback regulation.

In fact, the idea of noise decomposition in terms of different biophysical origins was first proposed by Paulsson in his pioneering work \cite{paulsson2004summing}. However, this work was focused on the decomposition of the local noise around the fixed point of the underlying biochemical reaction system, instead of the global noise of the entire probability distribution, by using the fluctuation-dissipation theorem, also called first-order van Kampen's expansion \cite{paulsson2005models}. In networks with no feedback, a decomposition of noise into the feedback-free noise $1/q\langle n\rangle$ and promoter switching noise $\eta_s$ can be found in \cite{pedraza2008effects, shahrezaei2008analytical}. In the present work, we obtain a noise decomposition in networks with feedback regulation, albeit in the regime of fast promoter switching. There are two major advantages of our decomposition formula \eqref{noise}. First, it can be applied to the situation when the nonlinearity of feedback regulation is very high. Second, all the three contributing terms in the decomposition formula can be estimated robustly from single-cell gene expression data.

In the regime of slow promoter switching, it is difficult to give an intrinsic definition of the promoter switching noise $\eta_s$ since the promoter switching rates $a_n$ and $b_n$ could be both nonlinear functions of the protein copy number $n$. In fact, an alternative definition of $\eta_s$ has been proposed in \cite{liu2016decomposition} with the aid of the thermodynamic limit of a piecewise-deterministic Markov process. By assuming linear feedback regulation and ignoring the mRNA kinetics, the authors decomposed the protein noise into the superposition of the protein birth-death noise, promoter switching noise, and correlation noise. Although their correlation noise is similar to our feedback coefficient (see the green curves in Figs. 2 and 3 of \cite{liu2016decomposition}), their protein birth-death noise is a constant independent of feedback regulation and thus is very different from our feedback-free noise.

Finally, we would like to point out that the lower bound of the protein noise in negative-feedback networks was first derived in \cite{lestas2010fundamental} by using concepts in information theory. However, this work is based on the diffusion approximation, with approximated Gaussian fluctuations, of the underlying discrete Markov model. A lower bound of the protein noise without diffusion approximation was derived recently in \cite{hilfinger2016constraints}. Our lower bound \eqref{lowerbound} is more explicit than the one obtained in \cite{hilfinger2016constraints} and is tighter in the regime of strong noise suppression.

Although we have shown how single-cell measurements may be used to reveal the feedback sign of a gene regulatory network, it is conceivable that in the near future, further advances in live-cell imaging with single-molecule resolution could allow the theory to be tested at the single-molecule level.

\section*{Methods}
The numerical simulations in Figs. \ref{Poisson} and \ref{simulations} are based on the Gillespie algorithm. The single-cell gene expression data of the synthetic gene circuit analyzed during this study are included in the published article \cite{shimoga2013synthetic}.

\section*{Acknowledgements}
We are grateful to the anonymous reviewers for their valuable comments and suggestions which helped us greatly in improving the quality of this paper. This work was supported by NIH grants MH102616, MH109665, K25AR063761 and also by NSFC 31671384 and 91329000. P. Xie acknowledges financial support form Eugene McDermott Graduate Fellowship.

\setlength{\bibsep}{5pt}
\small\bibliographystyle{nature}

\section*{Author Contributions}
C.J., M.C., and M.Q.Z. contributed to the formulation of the problem and development of the ideas. C.J. performed the mathematical derivations and numerical simulations. C.J. and P.X. contributed to the
single-cell data analysis. M.Q.Z. conceived the project and supervised the work. All authors participated in manuscript preparation, editing, and final approval.

\section*{Competing financial interests}
The authors declare no competing financial interests.

\end{document}